\documentclass{article}
\pdfoutput=1

\usepackage{arxiv}

\usepackage[utf8]{inputenc} 
\usepackage[T1]{fontenc}    
\usepackage{hyperref}       
\usepackage{url}            
\usepackage{booktabs}       
\usepackage{amsfonts}       
\usepackage{nicefrac}       
\usepackage{microtype}      
\usepackage{lipsum}
\usepackage{xcolor,soul}
\usepackage{graphicx}
\usepackage[numbers]{natbib}

\newcommand{\myul}[2][black]{\setulcolor{#1}\ul{#2}\setulcolor{black}}

\title{Legends: Folklore on Reddit}

\author{
\begin{tabular}{cc}
Caitrin Armstrong & Derek Ruths\\
\small{caitrin.armstrong@mail.mcgill.ca} &
\small{derek.ruths@mcgill.ca}
\end{tabular}\\
\\
School of Computer Science\\
McGill University
} 
\begin{document}
\maketitle

\begin{abstract}
In this paper we introduce Reddit legends, a collection of venerated old posts that have become famous on Reddit. To establish the utility of Reddit legends for both computational science/HCI and folkloristics, we investigate two main questions: (1) whether they can be considered folklore, i.e. if they have consistent form, cultural significance, and undergo spontaneous transmission, and (2) whether they can be studied in a systematic manner. Through several subtasks, including the creation of a typology, an analysis of references to Reddit legends, and an examination of some of the textual characteristics of referencing behaviour, we show that Reddit legends can indeed be considered as folklore and that they are amendable to systematic text-based approaches. We discuss how these results will enable future analyses of folklore on Reddit, including tracking subreddit-wide and individual-user behaviour, and the relationship of this behaviour to other cultural markers.
\end{abstract}


\section{Introduction}
Many parts of the internet place tremendous emphasis on new content.  Whether on Twitter, Facebook, Instagram, or Pinterest, platform design and conversational conventions focus on the newest posts, images, and trending hashtags.

Reddit shares in this culture of content generation, with one notable exception.  There is an aspect of Reddit behavior that revolves around venerated {\it old} posts.  True to Reddit's playful, self-referential attitude, these old posts will remain in the collective consciousness for years --- far outliving content of a similar age. They continue to be referenced and alluded to for so long that even the referencing users will not know the actual source post to which they are referring. We call these posts Reddit legends.  

These legends share surface similarity with the common conception of folklore. Folklore, broadly considered the shared artifacts of a group such as stories, can act to embody, reinforce and delineate the boundaries of a social group. In effect, folklore acts as a marker of belonging and a powerful purveyor of community beliefs and norms \cite{Bauman1971}.

As socialization and communities have moved increasingly online, a natural question, then, is what folklore exists among online communities? This is to say, can we identify digital cultural artifacts that are transmitted from one member to another and maintain special, stable significance within that online group?

To the extent that folklore acts as a cultural marker and carrier of beliefs and norms, the ability to detect and measure folkloric artifacts may provide a powerful device for identifying and characterizing salient communities in an otherwise fluid online space.

To this end, we investigate two main questions in this paper:
\begin{enumerate}
    \item{Are Reddit legends folklore?}
    \item{Can Reddit legends be studied systematically?}
\end{enumerate}

We first introduce folklore and existing scholarship on folklore in the digital age in order to highlight its significance, before considering culture on Reddit as well as previously identified phenomenon sharing similarities with folklore. We then address the first question using a set of simple yet thorough criteria for recognizing folklore, considering each of the criteria with respect to the Reddit legend phenomena in turn. We continue to build on our observations as we move to a computational approach that demonstrates the feasibility of studying Reddit legends through systematic means. We show how references to legends are widespread, and have recognizable patterns that can be leveraged in future computational tasks. 

Overall, we show that Reddit legends are folkloric in nature, and that they can be readily studied in a systematic manner. The primary analyses that we present, namely the creation of a typology and a methodology for the systematic extraction of referencing behaviour, will enable future exploration of specific hypotheses regarding folklore on Reddit. This will allow us to continue to unpack both online culture as well as contribute to our understanding of folklore on a broader scale. 

\section{Related Work}

\subsection{Folklore, from Traditional to Digital} Folkloristics literature highlights the significance of folklore within cultures. One simplified definition is that folklore is the ``vernacular expression'' \cite{blank2009folklore} of a particular group, although a common and precise definition of folklore has long eluded the field. \footnote{For a history, see \citet{oring1986folk}.} Most definitions, however, note the intertwined relationship between folklore and culture, often making a distinction between folk and "high" culture.  Folklore is also characterized by its dissemination within group boundaries through esoteric reference \cite{Bauman1971}.

Dundes discusses four main functions of folklore: amusement, cultural validation, education, and cultural stability \citet{dundes1965folklore}. Bauman, a foundational author in the field, extends this with folklore's significance for the social group, arguing that folklore is a function of shared identity and a representation of the collective character of a group. Folklore can also reinforce out-group boundaries as ``members of particular groups or social categories may exchange folklore with each other on the basis of shared identity, or with others, on the basis of differential identity'' \citet{Bauman1971}. According to Bauman folklore must satisfy three core criteria: (1) a consistent form (e.g. adhering to a motif or repeated structure), (2) stable cultural significance within the community, and (3) persistence by virtue of spontaneous transmission from community member to community member. It is these criteria we apply to to Reddit legends throughout this paper.  

In a contribution to ``Folklore and the Internet: Vernacular expression in a digital world'', Bronner distinguishes between a folk and an elite internet, examining the differences and similarities when considering ``tradition'' in the digital world \cite{blank2009folklore}. He notes that there are clearly folk groups on the internet, but is cautious about allowing the concept of ``tradition'' to apply frivolously to the fast-paced multi-actor behaviour of the internet, arguing that the thinning of boundaries also erodes the capability to assertively extract elements attributable to one clear group. He however acknowledges that a new understanding of folklore will need to take into account these differences. Other authors have also worked to remodel folklore scholarship for the digital age, arguing that internet phenomena such as Slender Man not only echo traditional formalizations of folklore, often invoking familiar behavioural customs and linguistic inflections, but are also key to the development of internet culture \citet{blank2012folk,blank2009folklore}. Several more recent pieces offer digital folklore as a lens through which we can gain understanding on public perception of contemporary issues. Through "hashtag ethnography", Lowthrop follows popular science hashtags on Twitter to examine the stories told about modern scientific advantages such as CRISPR, identifying widespread skepticism in the era of fake news \cite{lowthorp2018crisprfacts}. Digital folklore also finds a place in the political sphere with the advent of fake news and ``alternative facts'', as this newly-constructed folklore can reveal the values and aspirations of social groups. Fake news is often unverifiable just as traditional conceptions of folklore, but encourages the discussion of the relevant social issues nonetheless \cite{ellis2018fake,kitta2018alternative,mould2018doubt}. 

This literature offers compelling arguments for the role of folklore and folkloristics in our modern society, but to our knowledge lacks quantitative approaches - much less those that can be operationalized on a large scale. This paper showcases the exciting potential for interdisciplinary folklore research in the digital age.

\subsection{Reddit Culture and Community}

\subsubsection{Knowledge Collaboration and Sharing} Reddit is organized into subreddits focusing on a wide range of topics, each containing posts submitted by users. Redditors can then comment on these posts in threaded discussions. The word ``community'' is often used in discussions of subreddits, but no consensus has been reached as to whether all, or even some, subreddits \textit{are} communities. We find no works taking a firm decisive stance on this question, although a need has been expressed by several authors who broadly consider the question. Robards suggests we instead consider  subreddits as neo-tribes, characterized by ``by  fluidity,  occasional   gathering  and  dispersal'' \cite{Robards2018-zv}, while both \cite{kou2018knowledge} and \cite{kou2018you} consider certain subreddits as communities of practice, ``places where people with different levels of expertise can meet and collaborate''. Mills suggests we consider Reddit as a "collective intelligence approach to information overload" \cite{mills2011researching}. Using a more quantitative approach Olson and Neal combined network backbone extraction - extracting fundamental network components -  and community detection to show that subreddits can be viewed as part of a hierarchical map of topics, an observation shared by \cite{martin2017community2vec,Hessel2015-ny}. Reddit's subreddits are most readily described and organized by their topical content, rather than the interpersonal connections between their users. The role and significance of interpersonal connections on Reddit is so far unclear; information regarding the significance of interpersonal or group relationships on Reddit would allow for the application and use of social theories associated with different types of social organizations, particularity those relating to governance, lifespan, and formation. Legends likely form one piece of this puzzle: folklore has historically been a key component of creating and maintaining culture shared between individuals. 

When participating in subreddits, from the amusing to the serious, users contribute to the subreddit's collective knowledge. Work on motivations behind knowledge sharing has generally found that users are motivated by the uniqueness of their contributions as well as a feeling of a community bond \cite{ling2005using, ma2014knowledge, tausczik2014building}. Specifically on Reddit, surveyed users said they were drawn to the site because of its niche content \cite{newell2016user}. The visibility of content on Reddit - and thus what the userbase is most likely to interact with - is governed by functions between popularity expressed by user voting and time passed \cite{leavitt2017role}. Users who post successful content have engaged more on Reddit in the past \cite{liang2017knowledge} while more highly rated threads tend to be composed of heterogenous participants deeply discussing the topic at hand \cite{liang2017knowledge}. This offers a rich environment for the creation of folk culture, and an opportunity to study the relationships between participant type and folklore creation and referencing.

\subsubsection{Community Norms}
Users on Reddit often adhere to group norms, or ``regularities in attitudes and behavior that characterize a social group and differentiate it from other social groups'' \cite{hogg2006social}. There is extensive literature on the importance of norms for all communities, and a burgeoning field is investigating their role for online communities.

For example, both \citet{danescu2013no} and \citet{nguyen2011language} show that becoming a core member of a community means adopting community norms, and the extent of conformity to group norms reflects the commitment of a user to that group. Similarly, the distinctiveness of a community's language appears to explain some variance in user retention rates, and increasingly so when combined with the dynamicity (rate of language change) of an online community. Dynamicity also explains long-term user retention \cite{zhang2017community}. Comparing n-gram language models with topic models of several subreddits, Tran and Ostendorf find that style is a better indicator of community identity in a community classification experiment. In addition, they find a positive correlation between the community's reception to a contribution and the style similarity to the community to which that contribution was made \cite{tran2016characterizing}. Finally, loyal users engage with more niche content on subreddits, and employ language that signals collective identity \cite{hamilton2017loyalty}. Given that legend referencing is a particular form of community language that occurs in distinct stylistic forms, further investigations involving textual references may find similar effects. 

\subsubsection{Memes} 
Memes have emerged as an important component of online culture. While we find no direct references to Reddit legends in the literature, we find several describing online memes, including on Reddit. Nissenbaum and Shifman highlight the role of memes as the gatekeepers of online communities, where creation and knowledge of memes acts as cultural capital, allowing online users to gain social recognition from their detailed folk knowledge \citet{nissenbaum2017internet}. Literat and van den Berg  identify a similar, and much more self-aware (or ``meta'') phenomenon on the popular subreddit r/MemeEconomy, where users buy and sell memes in a mock stock exchange \citet{literat2019buy}. This type of meta-awareness and joy in self-determination is highly prevalent in Reddit culture \cite{singer2014evolution}.

A strong parallel can be drawn between Reddit legends and memes; they are very similar in both form and function. We find an important difference, however, in their transmission. A legend is a single object, usually textual. While a Redditor may forget the exact details of the original story its original content will almost always be accessible, and is itself the ``true artifact''. Memes, however, are more transient, we consider them to be ``a cultural unit defined by an atomic concept. A meme is identified by a name and a template and it can be implemented in different forms'' \cite{coscia2014average}. Online, this most often takes the form of an image, annotated with text. A meme has value at a specific point in time after which their novelty is lost \cite{literat2019buy}. Memes are meant to be reformulated and transformed, so while there are recognizable images they are constantly being remixed and taking on new meanings, whereas a legend is comparatively more stable. 

This is, however, a fine distinction, and we expect that many observations about memes will be equally relevant for legends. The study of legends on Reddit is highly complementary to these work and because of its textual nature enables a more readily-approachable quantitative investigation of their theses. 

\section{Dataset}


We have defined a Reddit legend as any non-informational post or comment that is frequently referenced or linked, despite being over 6 months old. It is important to disallow any old posts that are purely informational in nature.  Because Reddit is both a lively topical discussion platform and an informational resource, there are popular posts that remain in circulation simply because they are helpful as reference material. These posts or comments are often found in the form of collations of links to outside sources or expository guides, for example \href{https://www.reddit.com/r/AskReddit/comments/giolh/if_the_moderators_of_askreddit_were_to_put_a_list/}{\color{black} \myul[black]{this list}} of commonly asked questions on r/AskReddit. They may also take the form of explanations for bots - linked every time a bot such as the \href{https://www.reddit.com/r/AutoModerator/comments/q11pu/what_is_automoderator/}{\color{black} \myul[black]{widely-used moderator bot}} makes a post. Informational posts do not fulfill our conception of a legend: they do not carry additional meaning nor do they engender the same shared experience as legends.


We collect and analyze a dataset of Reddit legends for this paper, while recognizing that a significant challenge and contribution for the future will be to identify and publish more complete datasets of legends and references. To create our dataset we focused on the Reddit community itself as a guide. The subreddit \href{https://www.reddit.com/r/MuseumOfReddit/}{\color{black} \myul[black]{r/MuseumofReddit}} describes itself as ``a subreddit dedicated to cataloguing the posts and comments that will go down in Reddit history'', is heavily moderated to ensure duplicate and irrelevant content is removed, and has been active since 2013. We reviewed all 233 posts on this subreddit published before December 2016, where each post links to a possible legend, and discarded the following in order to create a consistent dataset:

\begin{itemize}
    \item Content posted by Reddit administrators. This content is often ``pinned'' as an announcement to the top of a post or the front page of Reddit, and therefore has a different potential for visibility than standard Reddit posts, presenting a problem for consistent data analysis, especially when considering referencing activity.
    \item Descriptions of famous Redditors unless their fame comes from a single piece of content: these Redditor's fame can come from thousands of posts, whereas we focus examining legends that are single units or events.
    \item Complex chains of events where there was no central or initial piece of content to reference, as for this type of legend URL references are highly dispersed, harder to conclusively collect, and textual references are often more complex, unnecessarily complicating the presentation of our results, which are intended to be an initial characterization of the phenomenon.
\end{itemize}

This resulted in a final list of 161 legends. We found that these legends are almost exclusively ``Reddit-wide'': without a focus so singular as to be applicable to only one subculture within Reddit. We recognize that our list is not exhaustive, however our set of legends is sufficient for an exploratory study. We also confirmed that our definition of a legend is in line with that of the community, none of the 233 legend candidates were primarily informational in nature, and all were at least 6 months old at the time of reviewing. 

The collected of this dataset, created in alignment with the definition of the community itself, allows us to investigate our two primary questions: are Reddit legends folklore, and can they be studied systematically? 

\section{Are Reddit Legends Folklore?}

Following Bauman, to be folklore, Reddit legend would have to satisfy three core criteria: (1) it must take on a consistent form (e.g. adhering to a motif or repeated structure), (2) it must have stable cultural significance within the community and (3) it must persist by virtue of spontaneous transmission from community member to community member \cite{Bauman1971}. We investigate each of these criteria in turn. 

\subsection{Do Reddit legends have consistent form?}

A preliminary characterization of the set of Reddit legends we collected suggests that a Reddit legend is, most often, a piece of well-written but not over-polished content, sincere in its message---as Redditors are militant against any attempt at planted or fake content \cite{bergstrom2011don}---and unrelated to outside social or political events.

Legends can take the form of links to media or Reddit-hosted text, and can additionally be divided into those that are standalone or those that unfold from a series of posts or comments. Most legends are textual in form. In almost all cases, the content was shared or made specifically for Reddit. This echoes Reddit's avowed support of original content.

To formalize these observations and examine whether these legends have a consistent form, we employed a multi-stage open coding process to create a typology to capture legend structure and content.

One author, a Reddit user for 8 years, first freely listed possible labels while reading through all legends. These labels were then examined for conceptual overlap and merged if necessary. The author then made a second pass to attach the finalized list of labels, and a third pass to ensure consistency. The final labels are not mutually exclusive: most legends are best described with two or three labels. We were able to easily assign a label to every legend in all rounds of annotation.


The labels were subdivided after the annotation process into 3 categories:
structural forms, affect types, and content types. On average, a legend is attached to 1.78 labels, with a maximum of 4 labels in 3 instances. We do not present a qualitative analysis of these labels due to the single-annotator paradigm and their highly subjective nature, but rather use them to introduce the reader to legend contents and show that they do have a consistent form.

\paragraph{Structural forms.} Legends assume one of two formats: single-post stories or series.  A \textbf{series} legend consists of a collection of posts and comments often with significant time delay between them. As a result, these legends are of additional entertainment value, as Redditors have time to speculate and anticipate the publication of a new update. In one instance, a Redditor first posts about how he is trying heroin (just once!) for the first time, updates as he proceeds to become addicted, eventually makes it to rehab and then \href{https://www.reddit.com/r/MuseumOfReddit/comments/68srty/spontaneoush_uses_heroin_gets_addicted_dies_gets/}{\color{black} \myul[black]{comes back}} with a positive life update 7 years later, having been clean 6 years.

The alternative to series structures --- and by far the most common structure on Reddit --- is the single-post narrative.  Because a significant amount of Reddit content takes this form, this structure isn't particularly noteworthy by itself.  That said, there are a number of legends that appear to have become legends solely on the merits of their writing.  We coded these as \textbf{strong narrative} content, characterized by well-written prose with a strong literary style, where Redditors have attempted to tell their peers a story, and tell it well. A \href{https://www.reddit.com/r/AskReddit/comments/1rgpdf/what_is_the_laziest_thing_youve_ever_done/cdnafqe/}{\color{black} \myul[black] {memorable example}} is that of a Redditor telling the story of a sea captain who gives the command to turn his vessel so that the sun is no longer in his eye; the text is written so that the reader is kept in suspense until the end. 

\paragraph{Affect types.} Certain legends have a strong emotive effect. \textbf{Heartwarming} content includes those showing goodwill towards others. An early \href{https://www.reddit.com/r/science/comments/6nz1k/got_six_weeks_try_the_hundred_push_ups_training/}{\color{black} \myul[black] {example}} includes a Redditor who corrects bad fitness advice given to one user, and then goes further, offering a comprehensive guide for the user on how to better himself. One of the most \href{https://www.reddit.com/r/getdisciplined/comments/1q96b5/i_just_dont_care_about_myself/cdah4af/}{\color{black} \myul[black]{famous legends}} is in a similar vein; a Redditor outlines a plan for personal improvement (beyond fitness), now summarized as the ``no zero days'' philosophy, reaching beyond Reddit. \footnote{A quick Google search of ``no zero days'' will reveal that Google has in fact indexed the post as such, despite no title being provided for Reddit comments. This highlights the popularity and appeal of these legends for many users.} In another of several examples with a real-world impact, Redditors \href{https://www.reddit.com/r/reddit.com/comments/dr6sy/kathleen_the_little_girl_with_huntingtons/}{\color{black} \myul[black]{banded together in 2010}} to give toys to a sick child. 
We also find legends meant to entertain. \textbf{Humorous} legends are either deliberately or accidentally funny, ranging from \href{https://www.reddit.com/r/AskReddit/comments/cfbkx/im_85_certain_that_there_is_an_adult_actress_in/c0s6bzw/?context=3}{\color{black} \myul[black]{puns}} to \href{https://www.reddit.com/r/AskReddit/comments/9ggfy/thag_see_problem_in_reddit/}{\color{black} \myul[black]{diatribes}} on Reddit.  \textbf{Surprising} content is either written to involve an element of surprise, is completely unexpected given the context, or describes an especially unusual event. Here examples include an \href{https://www.reddit.com/r/AskReddit/comments/cmwov/hey_reddit_what_tattoos_do_you_have/c0tpyls/}{\color{black} \myul[black]{increasingly inventive self-taken photo thread}} and a Redditor who \href{https://www.reddit.com/r/amiugly/comments/ldmyz/i_know_im_ugly_i_dont_even_know_why_im_here/c2rtxdy/}{\color{black} \myul[black] {photoshops}} another Redditor to improve his appearance by making him look more feminine. 

\paragraph{Content types.} Many legends stand out because of their content, itself. \textbf{Foolish} legends are those pieces of content that involved the Redditor doing something risky, unwise or just nonsensical. Foolish content differs from surprising content in that the subjects are  acting irrationally and describing a specific event. Among these are posts asking if you can put the \href{https://www.reddit.com/r/AskReddit/comments/iyj2n/so_if_you_pull_the_pin_on_a_grenade_can_you_put/}{\color{black} \myul[black]{put the pin back in a grenade}} (with no further updates), and (urgently!) \href{https://www.reddit.com/r/AskReddit/comments/gv1xy/urgent_need_to_know_right_fucking_now_is_it_safe/}{\color{black} \myul[black]{if octopi can drink beer}}.  Redditors often express admiration for the extreme acts found in these legends, reacting to them with a sense of awe or wonder. A number of legends can only be described as \textbf{gross}, inspiring a visceral feeling of \href{https://www.reddit.com/r/AskReddit/comments/t0ynr/throwaway_time_whats_your_secret_that_could/c4imcva/?context=3}{\color{black} \myul[black]{disgust}} or \href{https://www.reddit.com/r/pics/comments/jinex/shower_to_go/}{\color{black} \myul[black]{discomfort}} in the reader. This is typical of Reddit's orientation towards more extreme content, although only one is an explicit image, suggesting that widely shared content must appeal to a wide audience. Often but not always also gross, \textbf{sexual} content describes events sexual in nature. This category only includes those that are explicitly sex-related and does not include those that merely suggest at sexual content.

Finally, and of perhaps the most import from a folkloric perspective, we have a type of legend we chose to name \textbf{Reddit Power}. This type is the most prevalent, and describes content that exploits or celebrates Reddit as a social media website, usually involving Redditors working together in concert to create something humorous, amusing or worthwhile. Redditors fundraised to \href{https://www.reddit.com/r/IAmA/comments/fy6yz/51_hours_left_to_live/}{\color{black} \myul[black]{give a dying man his last vacation}} and \href{https://www.reddit.com/r/AskReddit/comments/a2xy8/would_anyone_be_interested_in_a_reddit_gift/}{\color{black} \myul[black]{started an international gift exchange}}. In another, \href{https://www.reddit.com/r/AskReddit/comments/msywj/what_is_the_most_mundane_thingsituation_that/}{\color{black} \myul[black]{two classmates met}} through a Reddit thread, discovering they were both posting in the same thread while in the same class. Reddit power legends subsume most but not all of the heartwarming content. This type of legend clearly holds significance for the community, as during the making of the legend Redditors came together, creating a shared experience, and thus an identity as Redditors.

\subsubsection{Summary}

Through this annotation task we found that Reddit legends do indeed have consistent and identifiable form. While several of the categories are highly subjective and thus difficult to examine in a multiple-annotator task i.e. it is difficult to agree on an exact definition of a humorous or foolish legend, the categories are easily recognizable and identifiable.

Although we initially prepared our typology of Reddit legends without reference to prior work on classifying folklore, we notice a correspondence between our chosen categories and Thompson's Motif-Index of Folklore \cite{thompson1989motif}, with similar categories including sex, taboo (gross),humour, marvels (heartwarming), and the wise and the foolish (foolish). These motifs cut across cultures, and Reddit is no different. 

\subsection{Do Reddit legends have function and significance?}

Here we examine the purpose that Reddit legends serve for the Reddit community, first through some further observations on the different category types and then through an examination of legends over time.

\subsubsection{The Function of Legend Types}

Our first observation is that there are certain legend types that need an additional "push" to become Reddit famous. Some collaborative legends, such as when Redditors reminded a user to do his homework, became famous only after repetition through a series of posts. When users come together in Reddit power posts, that action was usually displayed and gained momentum over a certain period of time. The mere fact that this can occur is motivational to the community and allows an event to be lifted into legendary status. Heartwarming content is never also gross or sexual; instead heartwarming posts are primarily Reddit Power posts as Redditors love to feel they are making a positive impact. The posts labelled gross, strong narrative or sex always have another label; they need another distinguishing feature to elevate them to legend status. Through this we can see that legends are not simple, everyday content, they must invoke some feeling or collective action beyond the ordinary. 
 
Given Reddit`s \href{https://www.vox.com/culture/2017/11/13/16624688/reddit-bans-incels-the-donald-controversy}{\color{black} \myul[black]{controversial reputation}} as a hotspot for hate speech and other bad behavior, it was entirely plausible that we would have need for ``hateful'', ``sexist'' or ``inappropriate'' labels.  However, none of the legends contain references to such content, save one, and that content only became legendary because of how the Reddit community came together to \textit{downvote} it, making this, in fact, heartwarming content. In addition, none of the legends make any sort of political claim or stance. While this does show neutrality in term of political function this is in itself significant, and opens the Reddit community up to those holding different political identities. 

It appears from this limited sample that collaborative efforts by the Reddit community tend away from the extremes Reddit has become infamous for. This is of course heavily influenced by our sample selected from only one subreddit, and future work is required. The repression of hate content in collective action is in line with previous work. The subreddit r/place was a collaborative social experiment where groups of Redditors fought to place pixels on a shared canvas to represent their group. While there was an early presence of hateful groups and associated imagery, by the final version 72 hours later all of these images had been wiped out, banishing hateful images from a work created collectively by Redditors. 

\subsubsection{Legend Evolution}

Reddit legends have changed as Reddit itself has changed. This may allow them to be identified as reflective of the community itself at the time. We offer some preliminary evidence for this claim in the following paragraphs.  

The earliest legend in our set is simply a comment ``\href{https://www.reddit.com/r/reddit.com/comments/6f7t/scientist_pours_molten_metal_into_ant_nests_w/c6far/?limit=1}{\color{black} \myul[black]{unfair to ants}}'' replying to a post linking a video of a scientist pouring molten metal into an ants' nest. This comment received 12 upvotes on reddit.com in 2006, before Reddit was subdivided into topical subreddits. Despite what is now a low number of upvotes ``unfair to ants'' was echoed by another redditor a few days later, and has remained in use by the savvy few even today, \href{https://www.reddit.com/r/OutOfTheLoop/comments/6a5fmb/whats_with_the_ants_why_are_they_mentioned_in_so/}{\color{black} \myul[black]{despite being occasionally confused with another legend}}.  Redditors have since developed a code of communication celebrating the creative, arcane, and witty exactly as exemplified in ``unfair to ants'' \cite{kasunic2018least}.

We observe that while gross or surprising content is found in all years, the ``mildest'' content is found only in the earliest years. A funny \href{https://www.reddit.com/r/funny/comments/eio6o/sometimes_when_i_get_a_text_from_a_wrong_number_i/}{\color{black} \myul[black]{wrong-number text conversation}} was amusing enough in 2010 to merit inclusion; such content is hardly remarkable on the internet of today. It is difficult to ascertain whether or not this is due to changing tastes or an over-saturation of duplicate content, as we observe that on the whole, legend content is rarely repeated; no two stories are alike, no photos can be judged as similar. The only exception to this is for ``Reddit Power'' posts which can involve similar acts of charity such as sending pizzas, indicating their symbolic and bond-forming significance for the community. 

\begin{figure}
    \includegraphics[scale=0.3]{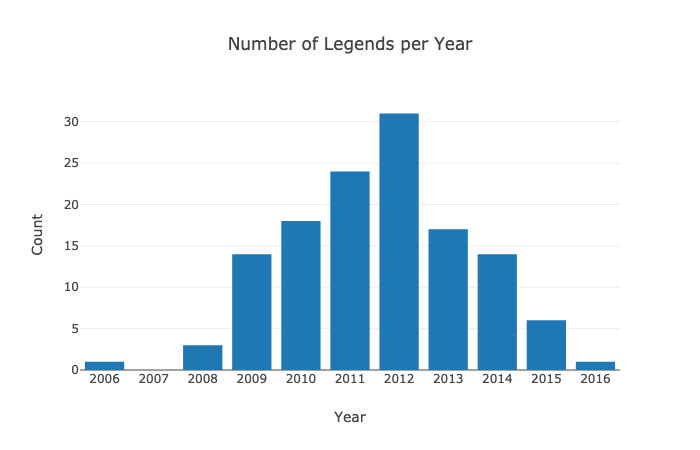}
    \caption{Number of legends per year. Note the steady rise until 2012, and then a sharp decrease. This perhaps corresponds with Reddit's growth and subsequent culture shift at that time, although the true cause is unknown.}
    \label{fig:number}
\end{figure}

Figure \ref{fig:number} shows the number of legends per year. There is a sharp increase in the number of legends created around 2012, and then a rapid decline, although there have been additional legends since our dataset cutoff point of December 2016. This is perhaps because content needs a longer time period to achieve legendary status, but could also be explained by the changing group size and traffic on the main (default) subreddits as well as increased member turnover, both factors contributing to participation rates as identified by Panek et al. \citet{panek2018effects}.

\subsubsection{Legend Type Interaction}
\begin{figure*}[h]
    \centering
    \includegraphics[width=0.9\textwidth]{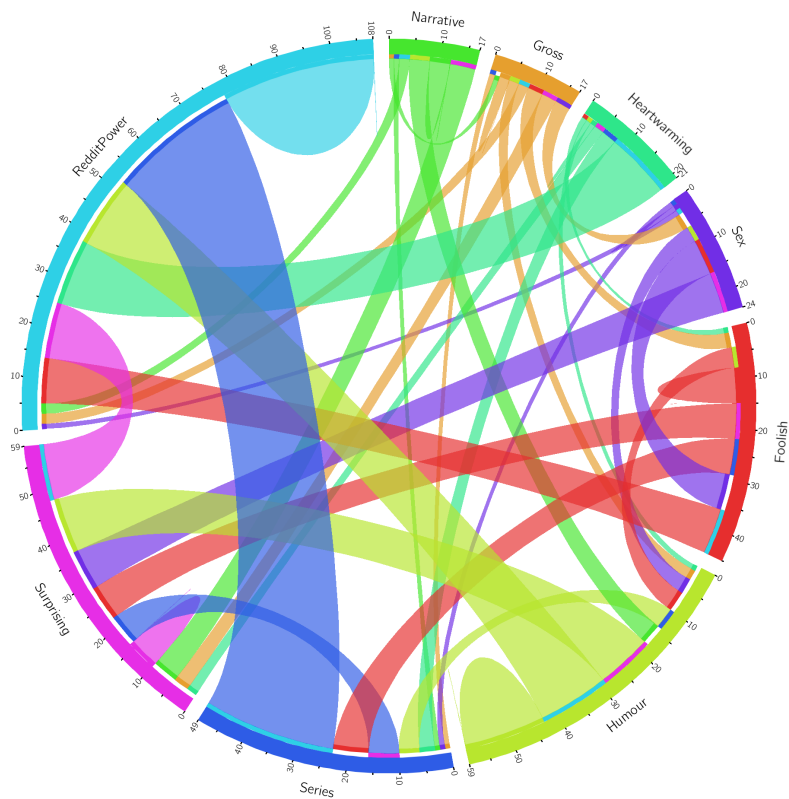}
    \caption{A circos plot showing the binary relationships between the legend labels. The width of the line illustrates the strength of the relationship between two labels, while the colours simply originate from the first label as a way of visual differentiation. The self-referential lines indicate when the legend type occurs alone with no other accompanying label.}
    \label{fig:circos}
\end{figure*}

One author, during the process of type extraction, also labelled each of the legends with the legend type(s) relevant to it. While we do not provide further quantitative analysis of type counts due to the single-annotator method, the the circos plot \cite{krzywinski2009circos} in Figure \ref{fig:circos} shows interactions are fairly disperse, there is no one clear partnership, save perhaps the strongest relationship between Reddit Power and Series: many collaborative legends, such as when Redditors reminded a user to do his homework, became famous only after repetition through a series of posts. Heartwarming content is never also gross or sexual; these are primarily Reddit Power posts as Redditors love to feel they are making a positive impact. The posts labelled gross, strong narrative or sex always have another label; they need another distinguishing feature to elevate them to legend status. 

\subsubsection{Summary}
We have provided a few preliminary observations of Reddit legend content. While the content of Reddit legends changes over time, it appears that they have a stable function within the community as a whole. We will provide further evidence, including some quantitative evidence, of the cultural significance of legends when examining legend referencing behaviour in the section on systematic approaches.

\subsection{Are Reddit legends transmitted? How are they transmitted?}

The third requirement is that legends be transmitted between Reddit users. While Reddit legends do differ from our standard conception of folklore in that their original source can almost always be readily identified and referenced, we find that other forms of transmission do exist, and appear to be the primary means of transmission of Reddit legends. Given that any given legend appears only once, and considering the sheer quantity of posts on Reddit, it is unlikely that any given user saw a legend the first time that it was posted. For example, in 2011,  Redditors overlooked 52\% of the most popular links the first time they were submitted \cite{gilbert2013widespread}. Therefore, references --- the mentioning or allusion to a legend --- are critical for disseminating knowledge of legends. Referencing behavior also acts to convey meaning, signal group membership and motivate the uninitiated to learn, just as in other forms of folklore. 

Most broadly, references to Reddit legends occur in two distinct forms: the textual reference and the URL reference. URL references simply consist of a link back to the original content, whereas a textual reference makes mention of a legend within the body of a Reddit comment. These textual references, often involving subtle wordplay, are hidden among the billions of Reddit comments. For example, the comment ``\$300 per day, huh? He can afford it" was made in reference to \href{https://www.reddit.com/r/AskReddit/comments/14uee5/how_long_would_you_let_someone_pee_on_you_for_300/}{\color{black} \myul[black]{a NSFW post}}  while "some people still think the puck is food" is making a reference to the post about  \href{http://www.reddit.com/r/hockey/comments/12edw2/update_would_it_mess_me_up_if_i_ate_a_puck/}{\color{black} \myul[black]{a goalie eating a hockey puck}}.

The URL references came from a dataset consisting of all Reddit comments from October 2007 through August 2017, compiled by Jason Baumgartner on pushshift.io. \footnote{ \href{https://files.pushshift.io/reddit/comments/}{\color{black} \myul[black]{The full set of Reddit data}}} We used regular expressions covering 25 unique linking forms to Reddit content, adapted from regular expressions created by user TheAppleFreak. \footnote{These were slightly modified from \href{https://www.reddit.com/r/AutoModerator/comments/4vapin/need_to_filter_links_to_reddit_threads_heres_the/}{\color{black} \myul[black]{the originals}}} We found 58,574,317 links to valid Reddit content. 92,274 of these were references to one of our 161 legends. There was only one possible link per legend, however for legends in the form of comments we included links referencing any content within the same post, in recognition of the fact that many comment-based legends occurred as a series of comments.

On average, each non-series legend \footnote{We excluded these (35 in total) from this analysis as there many separate links for each series that were not collected in our original reference set.} received 671.4 references ($\sigma = 1167$). The most popular legend was referenced 7969 times, this NSFW legend is most frequently referred to as "broken arms". One legend, the original comment about "unfair to ants", was referenced directly only 4 times. However, a Reddit search for this phrase will reveal many more textual references as the original source was apparently quickly lost. The subreddit containing the most referencing comments was r/AskReddit with 36189 references, while the second, r/AdviceAnimals contained only 6091 in comparison. It is worth noting that the top referencing subreddits were all previous default subreddits \footnote{\href{https://www.reddit.com/r/announcements/comments/6eh6ga/reddits_new_signup_experience/}{\color{black} \myul[black]{These were formally removed in May 2017}}}, which represent some of the most trafficked and well-recognized subreddits generally visited by many members of the Reddit community. Only \href{https://www.reddit.com/r/hiphopheads/comments/1o3vse/daily_discussion_thread_10092013/ccom8d5/}{\color{black} \myul[black]{one legend}} did not have the majority of its references coming from these "default" subreddits; it is likely that there are many more legends specific to individual subreddits but that these were not included on the site-wide r/Museumofreddit. We did note an apparent relationship between the legend type (i.e. funny) and the selection of subreddits containing references to those legends. While we do not include an analysis of legend content and referencing subreddits due to space and annotation validation concerns, future work should consider this relationship.

\subsubsection{Summary}
Overall, it is clear that Reddit legends are transmitted between users, through both text and URL references. Future work should consider trends across subreddits and users.

\section{Can Reddit Legends be Studied Systematically?}

To prove useful to both the HCI community and folklorists wishing to bring qualitative methods to their study, it is necessary that Reddit legends can be studied through systematic (computational) means. 

To be able to study the dynamics of legend creation and legend sharing, we must be able to identify all or many of the references that are made to a Reddit legend. As we have seen, these references occur in two main forms: direct URL links and textual references. While we found 92,274 URLS referencing our identified legends, it is a reasonable assumption that many more textual references also exist. It is however difficult to identify these references in any systematic manner given the presence of billions of other comments, and the complexity of the referencing phenomenon as illustrated above. If the target is a training set of at least several hundred reference examples per legend, it is unreasonable to expect that these references be found by combing through Reddit content. Using keyword search will bias the available samples, although a boostrapping keyword strategy may prove useful. 

To overcome these limitations we propose an alternate sampling strategy for the creation of such a training set and demonstrate its feasibility. 

\begin{figure}[h]
    \includegraphics[width=0.3\textwidth]{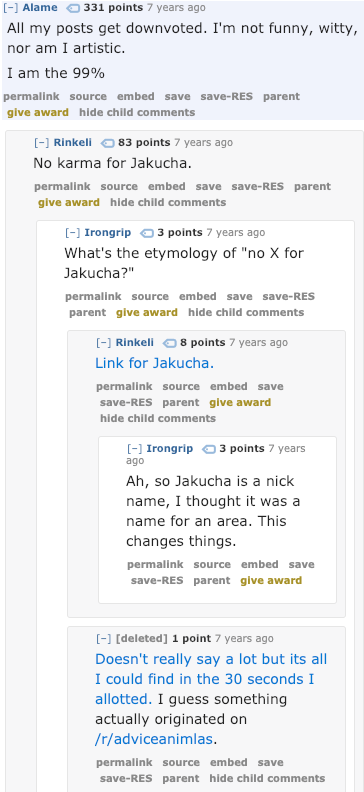}
    \centering
    \caption{A typical conversation about a Reddit legend: one user is prompted to make a reference, another is confused and asks for an explanation, and then is provided numerous helpful explanations.}
    \label{fig:jakucha}
\end{figure}

We take advantage of a convenient common conversation motif: we observed that it is not uncommon for one Redditor to make a textual reference to a legend, soon after another Redditor will ask about that reference (understanding that he has missed something significant but not knowing what it is), and then a third Redditor will reply, providing a URL link to the original legend. Figure \ref{fig:jakucha} provides an example. This pattern provided a direct method for identifying textual references for all the legends we considered in this study.  Our process began with collecting all comments containing URL references to our set of Reddit legends.  We then used multiple annotators to manually search for the textual references in the preceding comments. It is worth noting that this is, almost certainly, a biased dataset: it will favor more vague/confusing textual references, less well-known legends, and certain kinds of user interactions. For this reason, we use our data only to provide an exploratory characterization of referencing patterns. 

Of the full set of legends, our analysis considered exclusively those that are textual (as opposed to those with an external link or image) because they were amenable to natural language processing. This resulted in a candidate list of 99 text-based legends and 47,906 referencing comments to these legends. Out of all these, 14,756 were top-level comments (direct reply to a post), while 33,150 had at least one parent. 

To prepare for the annotation task, we re-created the context surrounding our dataset of referencing comments. For all of the non top-level references to the text-based legends we walked up the tree of ancestors to a maximum of 3 ancestors and collected the text body of each comment. We then reformatted the texts in order to emulate a Reddit thread, which contains successive indented comments that are ordered by time, with the most recent at the bottom. We randomly selected 40 legends and from each of these legends 80 random comment threads to use for the annotation task.

\subsubsection{Annotation Task}
We recruited Redditors with at least one year of experience from our university subreddit, under the approval of our university ethics board. We ensured our  annotators had an understanding of Reddit behaviour as we wanted to achieve the highest recall possible. Each annotator candidate, recruited from our university subreddit, was asked to answer 7 out of 10 questions correctly on a quiz before being allowed to participate. This quiz was written by the first author and 3 other members of our research group, all Redditors for over 5 years, average age approximately 25. The short-answer questions tested the candidates' knowledge of common Reddit legends, customs, and norms. It also ensured that they were already well-exposed to the often highly-offensive and ``not safe for work'' content found on Reddit. On the first round of selection almost all annotator candidates (approximately 80\%) who filled out the quiz failed with only 2-3 of questions correct despite claiming to be frequent Redditors. Informal polling suggested they had an average age of  approximately 20 years and 2 years of Reddit experience. The candidates suggested that we amend the quiz to include more recent legends. After doing so nearly all annotator candidates - all new non-repeat candidates -  successfully passed the quiz. Most of our research group were not familiar with the legends, but recognized them nonetheless as valid, having the implicit hallmarks of legends. This suggests that recall and knowledge of the most memorable and significant legends varies according to the demographic of the user. This does not imply a lack of significance of legends but rather the opposite - that they are so significant to certain demographics that others are not as readily communicated or recognized. 

The annotation task proceeded as follows: using the brat annotation software \cite{stenetorp2012brat} we asked our annotators to highlight and label the text in each reconstructed thread that contained either a direct reference, indirect reference, question or an extra category titled ``other relevant text''. A direct reference was explained to be those references that contained legend label, effectively naming it, while indirect references contained no label but rather made an allusion towards the legend. A prototypical example was given for direct as ``the guy who ate a puck'' and indirect as ``wasn't there someone who ate a puck''. We included the other relevant text category to allow for the possibility of having excluded an important alternate classification. These categories were chosen so as to allow us a rough typology of references in order to examine their significance for future tasks. As many selections as necessary were allowed per thread. Pairs of annotators worked separately on randomly assigned threads selected from 8 legends without replacement. Two pairs of annotators covered the same 8 legends, but each pair worked on a different set of threads, so as to increase coverage and fidelity per legend annotated. Before beginning the task each annotator was given the link to the original content and asked to review it so that they were familiar with the content.
\subsubsection{Annotator Agreement}
Thirteen annotators participated, with one annotator working for 2 hours instead of 1. One annotator was removed due to low task participation. 20 legends had greater than 20 comment threads annotated and were retained for the analyses below. Overall, these generated a total of 519 annotated threads, 314 of which were dual-annotated. The average Cohen's kappa among annotators sharing at least 80 comment thread lines (N=5) was 0.62, with a variance of 0.03, min 0.33, max 0.81. Annotators were considered to be in agreement when they identified the same line of the comment thread as containing a reference. Each identification by an annotator was considered to be a possible source of agreement or disagreement, but no double-counting occurred. A post-task manual analysis of 200 annotated threads revealed that the disagreement between users was not one of false positives, but rather of omission of true positives. In these cases, they most frequently omitted indirect references, missed subtle information, or disagreed on the nature of a deliberate reference. In less than 5 of 200 instances, a user annotated irrelevant text. We find, therefore, that while annotator agreement is not high, the double annotations increased the recall of our dataset.

This points to a fundamental difficulty within our task: legend references can be quite subtle and even occasionally coincidental (i.e. it is unclear from the comment if a user actually meant to reference a legend). The ability to automatically identify legend references with high fidelity is an intriguing area of future research and this annotated dataset and additional data generated in the same way will be an indispensable tool. Annotator agreement could be improved in the future once the legend referencing phenomenon is more fully understood. The distinction between direct and indirect references is important to make, as it is likely that direct references will be much easier to identify, or at least that these two categories will have very different signatures. This information will be important for future classification tasks. It remains, however, that we have shown the feasibility of such a process. 

\subsection{Prevalence and Type of Textual Reference}

Within the retained threads, there were 529 ``indirect'' references labelled and 815 ``direct'', with 287 phrases labelled as ``question''. We removed ``other relevant text'' annotations as this category was seldom used (42 instances) and without any consistent pattern emerging, aside from its use to indicate references to other non-target legends. 

\subsection{Reference Relation to Original Legend}

Given that references take the place of what was previously an oral retelling of a legend, it is pertinent to examine the relationship between the legend and reference texts. We provide an initial characterization in two ways: the number of terms shared between references and legends and the terms most often found across references. 

\begin{figure}[htbp]
    \includegraphics[scale=0.35
    ]{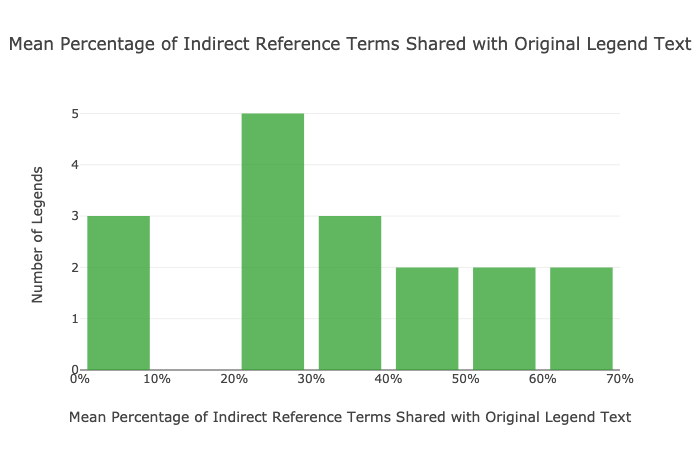}
    \caption{Histogram of the mean percentage of terms shared with the original legend text for each collection of indirect references to a Reddit legend. To illustrate: There were five legends whose references shared, on average, between 20-30\% of their terms with the original text.}
    \label{fig:refterms}
\end{figure}

We tokenized both legend and reference texts and extracted unigrams (removing stopwords). We then calculated the number of reference unigrams that also occurred in the original legend text. We present the mean percentage of shared unigrams in figure \ref{fig:refterms}: values vary per legend from 0 to 70\%, a high percentage of shared terms. Importantly, we find no significant correlation of this result with the length of the legend text using Pearson's correlation coefficient ($r$ =0.33, $p$ = 0.20). However, we do observe that the smallest shared percentages occur with those legends whose legend status was unrelated to the original content, (e.g. the \href{https://www.reddit.com/r/pics/comments/2x9zvi/president_obama_thanks_redditors_for_their_help/coyaibj/}{\color{black} \myul[black]{time u/jstrydor wrote their own name wrong}}), while those that have the highest percentage are \href{https://www.reddit.com/r/AskReddit/comments/1sjt6n/if_you_could_only_post_the_same_one_sentence_to/cdyaopn/}{\color{black} \myul[black]{highly memorable texts}} that are often repeated verbatim or rephrased slightly. Future analyses should examine the relationship between text relevance and memorability and legend popularity.

\begin{figure}[htbp]
    \includegraphics[scale=0.35]{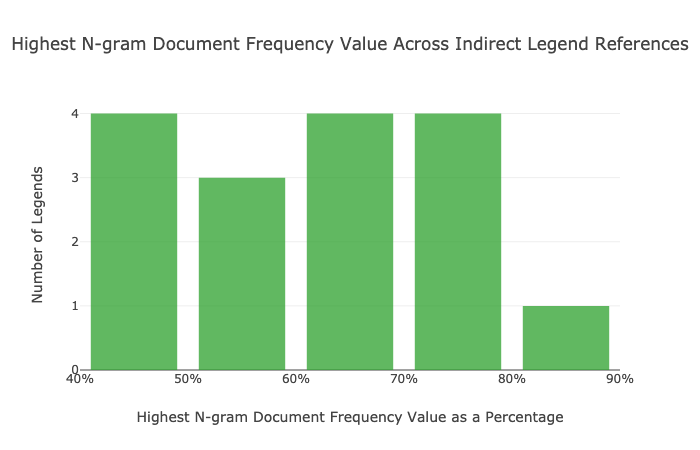}
    \caption{Histogram of the highest document frequency value for each collection of indirect references to a Reddit legend. For illustration, 4 legends have a term that is shared across 60-70\% of all references to that legend.}
    \label{fig:docfreq}
\end{figure}

We also examined the terms most commonly found across legend references. For every legend and reference type having at least 10 annotated comments, we calculated the document frequency value of every unigram, bigram, and trigram, with stopwords removed from the unigram set. We then calculated, for every n-gram, the proportion of documents in which it occurred. We present a histogram of the highest observed value for every collection of indirect references to a legend  (figure \ref{fig:docfreq}). Many of these values are quite high, going up to 90\% in some cases, indicating the presence of terms that occur frequently across all documents. Upon examination, many of these terms are proper nouns that serve as a unique identifier of a legend; e.g. a Reddit search of ``Jakucha'' or ``Wadsworth'' will result in a page of legend references. Of those legends that are not quite so lucky as to have a unique identifier, we find that their references contain longer descriptive n-grams with lower document frequency values, such as ``banned in Russia''. This offers evidence that there are concrete patterns that can be extracted from legend referencing behaviour, which relates back to our requirement that the Reddit legend phenomenon displays consistent structure and form: we would not have observed such high co-occurence values without consistent patterns of referencing behaviour. 


\subsection{Why is a reference used?}


The collection of these references also allows us to further add to our knowledge about the function and significance of Reddit legends.

From the set of annotated references, we randomly selected 80 of those references with at least 2 non-top-level and non-deleted comments above them, and printed out the entire comment chain. Using a two-pass open-coding approach, one author identified the common reasons for referencing a legend, discovering 4 main categories. These categories include:

\begin{itemize}
\item \textbf{Similar Situation}: the description in the instigating comment is similar enough to the events described by the legend
\item \textbf{Similar Wording}: the phrasing or words in the instigating comment are similar enough to the legend text or to the common style of legend reference. These are generally wordplay or dependent on just a few keywords, rather than on an entire description of a situation. 
\item \textbf{Comparison}: either another Reddit legend is referenced, and thus the legend is brought up comparatively, or someone brings up the legend to beat a boast of ``this is the best legend ever''.
\item \textbf{User Implicated}: occurs when a user implicated in a legend is referenced by name or when they make a post.
\end{itemize}

While we chose a single most-appropriate category for every referencing comment chain, the difference between ``similar wording'' and ``similar situation'' is subtle and that these categories can often overlap, as similar situations beget similar wording.

These referencing categories indicate that referencing behaviour occurs as a method of humorous interpersonal and person-group interaction, with users employing legends to showcase their knowledge and relevance, finding unique ways to integrate a little bit of Reddit culture into the conversation. This appears to be congruent with the idea of cultural capital, as users share to establish their knowledge and significance within the community, although the true mechanism of action is unclear given that Redditors do not necessarily establish interpersonal relationships. 

We also note that almost of the references carried a sense of ``play''; while serious discussions do occur on Reddit, legend referencing appears to not be one of those occasions. While the Reddit legends and references in our dataset are varied, they share common characteristics and are on the whole playful and accessible for all Redditors. We find similarity to the ``logic of the lulz'' or the archetype of the trickster used to describe meme, hacker, and troll culture on both Reddit and 4chan, where users use sharp offensive humour and irony to establish social status \cite{milner2013fcj,coleman2014hacker}. In this context, however, the ``lulz'' do not come at the expense of the marginalized groups, and are generally palatable to most Redditors. 

Finally, we observed no social consequences for question asking, unlike users on 4chan who are subject to social disgrace for the misuse of a culturally relevant meme or structural style of text \cite{nissenbaum2017internet}. Again, this offers evidence of consistency of structure in legends and legend referencing behaviour.

\section{Limitations}

The primary limitations faced in this study are the single-annotator design for 2 annotation tasks, of which the legend typology is the most central. We ultimately chose this design for its descriptive simplicity in the face of extremely subjective data. The labels, while mutually understandable categories, were not amenable to conflict resolution, such as what constituted ``humour'' or ``grossness''. A single annotator allowed for a more consistent assignment of labels. The other task was singly-annotated for the same reason: highly subjective categories intended not as concrete category creation but as sense-making to suggest directions for more concerted research. 

Other limitations include the choice of sample. The use of one subreddit as a single source for legend collection undoubtedly causes bias in unidentifiable ways. This is however a choice that allows for high precision, if not necessarily high recall; we feel certain that we collected content almost universally agreed to be Reddit legends, given that this is the most popular subreddit for cataloguing them. Finally, in regards to the reference annotation task we selected only textual references that were followed by a URL reference, there are surely many more that do not share this structural form. However, given that identifying legend references out of billions of Reddit comments is a difficult unsolved computational problem unaddressed by current techniques, we feel that this is a crucial first step towards understanding Reddit legend referencing behaviour. 

\section{Future Work}

Folklore serves as a valuable tool for mapping culture and revealing social issues and values. Future work can both exploit this capacity and expand its potential.  

The ability to extract textual references would be an invaluable tool for mapping this phenomenon at a cultural and user level. Our annotation task showed that a training set can be formed for future extraction tasks. Such an ability would allow us to explore if subcultures, such as subreddits or sets of subreddits, have similarly readily-identifiable folkloric structures. If they do, which we believe is likely, references to these legends could be used to delineate cultural boundaries. User personality and preferences could be revealed, for example if  users prefer content that was originally posted when they join a site. Such an analysis would support work such as that of Danescu et al. \citet{danescu2013no}, who predict user's lifecycles based on their speed of adaptation to changing community norms. We may also be able to use legends to track community identification over fluid online spaces, such as user migration from one subreddit to the next, an occurrence that has been previously identified in several spaces including on Reddit and fandom groups \cite{dym2018generations,newell2016user}.

If legends are indeed a type of cultural capital, acting as a method of gaining social status, then we could expect to see a type of users who ``deals'' in legend references. An analysis of user influence  such as this would enable identification of key actors vital to a healthy community.

\section{Conclusion}

We showed that Reddit legends are folkloric in nature, as they have consistent form, cultural significance and are persisted through spontaneous transmission between group members. They can also be readily studied in a systematic manner by extracting references and examining the characteristics of legend and reference text. The primary analyses that we present, namely the creation of a typology and a methodology for the systematic extraction of referencing behaviour, will enable future exploration of specific hypotheses regarding folklore on Reddit, including those related to subreddit-wide and user-specific behaviour. Both folkloristics and previous work in computational social science offer exciting lenses through which we can view the Reddit legend phenomenon.

\bibliographystyle{unsrtnat}
\bibliography{references}

\end{document}